\documentclass[a4paper,11pt]{article}
\usepackage{aaskaiid}
\usepackage{orcidlink}

\title{Galactic Centre Pulsars with the SKAO}
\ShortTitle{Galactic Centre Pulsars with the SKAO}

\author[1,2]{F.~Abbate\orcidlink{0000-0002-9791-7661}}
\ShortName{Abbate et al.} % shortened name list for header 
\author[1]{A.~Carleo\orcidlink{{0000-0001-9929-2370}}}
\author[3]{S.~Chatterjee\orcidlink{0000-0002-2878-1502}}
\author[3]{J.~Cordes\orcidlink{0000-0002-4049-1882}}
\author[4]{P.~B.~Demorest\orcidlink{0000-0002-6664-965X}}
\author[2]{G.~Desvignes\orcidlink{0000-0003-3922-4055}}
\author[5,2]{R.~P.~Eatough\orcidlink{0000-0001-6196-4135}}
\author[6]{E.~Hackmann\orcidlink{0000-0001-7456-216X}}
\author[7,8]{Z.~Hu\orcidlink{0000-0002-3081-0659}}
\author[2]{M.~Kramer\orcidlink{0000-0002-4175-2271}}
\author[9]{J.~Lazio}
\author[7,5,10,11]{K.~J.~Lee\orcidlink{0000-0002-1435-0883}}
\author[12,2,13]{K.~Liu\orcidlink{0000-0002-2953-7376}}
\author[2]{I.~Rammala-Zitha\orcidlink{0000-0002-0224-6579}}
\author[4]{S.~M.~Ransom\orcidlink{0000-0001-5799-9714}}
\author[14]{G.~Saowanit}
\author[8,5,2]{L.~Shao\orcidlink{0000-0002-1334-8853}}
\author[15,2]{P.~Torne\orcidlink{0000-0001-8700-6058}}
\author[2]{R.~Wharton\orcidlink{0000-0002-7416-5209}}
\author[16,2]{J.~Wongphechauxsorn\orcidlink{0000-0002-7730-4956}}
\author[5,17]{W.~Zhu\orcidlink{0000-0001-5105-4058}}
\author[]{The SKAO Pulsar Science Working Group}

\affiliation[1]{INAF-Osservatorio Astronomico di Cagliari, Via Della Scienza 5, I-09047 Selargius, Italy}
\emailAdd{federico.abbate@inaf.it}
\affiliation[2]{Max-Planck-Institut für Radioastronomie, Auf dem Hügel 69, D-53121 Bonn, Germany}
\affiliation[3]{Cornell Center for Astrophysics and Planetary Science and Department of Astronomy, Cornell University, Ithaca, NY 14853, USA}
\affiliation[4]{National Radio Astronomy Observatory, 1003 Lopezville Rd., Socorro, NM 87801, USA}
\affiliation[5]{National Astronomical Observatories, Chinese Academy of Sciences, 20A Datun Road, Chaoyang District, Beijing 100101, P.R. China}
\affiliation[6]{University of Bremen, Center of Applied Space Technology and Microgravity (ZARM), 28359 Bremen, Germany}
\affiliation[7]{Department of Astronomy, School of Physics, Peking University, Beijing 100871, China}
\affiliation[8]{Kavli Institute for Astronomy and Astrophysics, Peking University, Beijing 100871, China}
\affiliation[9]{Jet Propulsion Laboratory, California Institute of Technology, 4800 Oak Grove Drive, Pasadena, CA 91109, USA}
\affiliation[10]{Yunnan Astronomical Observatories, Chinese Academy of Sciences, Kunming 650216, Yunnan, P. R. China}
\affiliation[11]{Beijing Laser Acceleration Innovation Center, Huairou, Beijing 101400, P. R. China}
\affiliation[12]{Shanghai Astronomical Observatory, Chinese Academy of Sciences, Shanghai 200030, P. R. China}
\affiliation[13]{State Key Laboratory of Radio Astronomy and Technology, A20 Datun Road, Chaoyang District, Beijing, 100101, P. R. China}
\affiliation[14]{Department of Physics, Faculty of Science, Kasetsart University, Bangkok 10900, Thailand}
\affiliation[15]{Institut de Radioastronomie Millim\'etrique, Avda. Divina Pastora 7, Local 20, E-18012 Granada, Spain} 
\affiliation[16]{Julius-Maximilians-Universität Würzburg, Institut für Theoretische Physik und Astrophysik, Lehrstuhl für Astronomie, Emil-Fischer-Straße 31, D-97074 Würzburg, Germany}
\affiliation[17]{Institute for Frontier in Astronomy and Astrophysics, Beijing Normal University, Beijing 102206, People’s Republic of China}

\abstract{The detection of a pulsar closely orbiting our Galaxy's supermassive black hole - Sagittarius~A* - 
is one of the ultimate prizes in pulsar astrophysics. The relativistic effects expected in such a system could far exceed those currently observable in compact binaries such as double neutron stars and pulsar white dwarfs. In addition, pulsars offer the opportunity to study the magneto-ionic properties of Earth's nearest galactic nucleus in  unprecedented detail. For these reasons, and more, a multitude of pulsar searches of the Galactic Centre have been undertaken, with the outcome of just seven pulsar detections within a projected distance of 100\,pc from Sagittarius~A*. It is currently understood that a larger underlying population likely exists, but it is not until observations with the SKA have started that this population can be revealed. In this chapter, we look at important updates since the publication of the last SKAO science book and offer a focused view of observing strategies and likely outcomes with the updated SKAO design.}

\begin{document}

\newcommand{\actaa}{Acta Astron.} % Acta Astronomica
\newcommand{\araa}{ARA\&A} % Annual Review of Astron and Astrophys
\newcommand{\aar}{A\&ARv} % Astrononmy \& Astrophysics Review
\newcommand{\aapr}{A\&ARv} % Astronomy\&Astrophysics Reviews
\newcommand{\ab}{Astrobiol.} % Astrobiology
\newcommand{\aj}{AJ} % Astronomical Journal
\newcommand{\apj}{ApJ} % Astrophysical Journal
\newcommand{\apjl}{ApJL} % Astrophysical Journal, Letters
\newcommand{\apjs}{ApJSS} % Astrophysical Journal, Supplement
\newcommand{\ao}{Appl. Opt.} % Applied Optics
\newcommand{\apss}{Astro. \& Space Sci.} % Astrophysics and Space Science
\newcommand{\aap}{A\&A} % Astronomy and Astrophysics
\newcommand{\aaps}{A\&AS.} % Astronomy and Astrophysics, Supplement
\newcommand{\baas}{Bull. Am. Astron. Soc.} % Bulletin of the AAS
\newcommand{\caa}{Chinese A\&A} % Chinese Astronomy and Astrophysics
\newcommand{\cjaa}{Chinese J. A\&A} % Chinese Journal of Astronomy and Astrophysics
\newcommand{\cqg}{Class. Quantum Gravity} % Classical and Quantum Gravity
\newcommand{\gal}{Galaxies} % Galaxies
\newcommand{\gca}{Geo. Cosmo. Acta} % Geochimica Cosmochimica Acta
\newcommand{\icarus}{Icarus} % Icarus
\newcommand{\jcap}{JCAP} % Journal of Cosmology and Astroparticle Physics
\newcommand{\jgr}{J. Geophys. Res.} % Journal of Geophysics Research
\newcommand{\jgrp}{J. Geophys. Res. Planets} % Journal of Geophysics Research: Planets
\newcommand{\jqsrt}{J. Quant. Spectrosc. Radiat. Transf.} % Journal of Quantitiative Spectroscopy and Radiative Transfer
\newcommand{\memsai}{Mem. SAIt} % Mem. Societa Astronomica Italiana
\newcommand{\mnras}{MNRAS} % Monthly Notices of the RAS
\newcommand{\nat}{Nature} % Nature
\newcommand{\nastro}{Nat. Astron.} % Nature Astronomy
\newcommand{\ncomms}{Nat. Commun.} % Nature Communications
\newcommand{\nphys}{Nat. Phys.} % Nature Physics
\newcommand{\na}{New Astron.} % New Astronomy
\newcommand{\nar}{New Astron. Rev.} % New Astronomy Review
\newcommand{\physrep}{Phys. Rep.} % Physics Reports
\newcommand{\pra}{Phys. Rev. A} % Physical Review A: General Physics
\newcommand{\prb}{Phys. Rev. B} % Physical Review B: Solid State
\newcommand{\prc}{Phys. Rev. C} % Physical Review C
\newcommand{\prd}{Phys. Rev. D} % Physical Review D
\newcommand{\pre}{Phys. Rev. E} % Physical Review E
\newcommand{\prx}{Phys. Rev. X} % Physical Review X
\newcommand{\prl}{Phys. Rev. Let.} % Physical Review Letters
\newcommand{\psj}{Planet. Sci. J.} % Planetary Science Journal
\newcommand{\planss}{Planet. Space Sci.} % Planetary Space Science
\newcommand{\pnas}{Proc. Natl Acad. Sci. USA} % Proceedings of the US National Academy of Sciences
\newcommand{\procspie}{Proc. SPIE} % Proceedings of the SPIE
\newcommand{\pasa}{PASA} % Publications of the Astron.  Soc. of Australia
\newcommand{\pasj}{PASJ} % Publications of the Astron.  Soc. of Japan 
\newcommand{\pasp}{PASP} % Publications of the Astron.  Soc. of the Pacific
\newcommand{\rmxaa}{RMXAA} % Revista Mexicana de Astronomia y Astrofisica
\newcommand{\sci}{Science} % Science
\newcommand{\sciadv}{Sci. Adv.} % Science Advances
\newcommand{\solphys}{Sol. Phys.} % Solar Physics
\newcommand{\sovast}{Soviet Ast.} % Soviet Astronomy
\newcommand{\ssr}{Space Sci. Rev.} % Space Science Reviews
\newcommand{\uni}{Universe} % Universe

\maketitle

\section{Introduction}

A decade has passed since the publication of the last report on the prospects of observing pulsars in the Galactic Centre (GC) with the SKAO \citep{Eatough+2015}. Over this time, the interest for the GC has grown and reached a high point thanks to the Nobel Prize in Physics 2020 awarded to Roger Penrose, Reinhard Genzel and Andrea Ghez in part for the discovery of the supermassive compact object at the centre of our Galaxy. In view of all the novel science results surrounding the GC, we discuss the updates regarding pulsar science and the prospects of discoveries using the current re-design of the SKAO.

A wide range of independent astronomical observations - using entirely different methods and at multiple wavelengths - now provide a wealth of evidence indicating that Sagittarius~A* (Sgr~A*) - the ultra compact radio source in the GC - is indeed a supermassive black hole.  

One method involves high precision astrometry in the near infrared. Great effort has been spent on the precise tracking of the S0-2 star that orbits around Sgr A* every $\sim 15$ years \citep{grvty+18,grav20_precS2,Do2019}. These efforts led to the detection of several relativistic effects like gravitational redshift and Schwarzschild precession that are compatible with the predictions of General Relativity (GR). The detection in infrared of a hot spot orbiting around Sgr A* at about 6-10 gravitational radii and moving at almost 30 percent the speed of light further confirmed these results \citep{grav18}. 

A different experiment using novel interferometric techniques at millimeter wavelengths redefined the concept of black hole imaging. The Event Horizon Telescope (EHT) Collaboration was able to create images of the supermassive black holes M87* and Sgr A* detailed enough to see the shadow cast by the black holes on the surrounding plasma \citep{eht19a,eht22a}. When compared to synthetic images based on general relativistic magnetohydrodynamic simulations, the EHT results provide a new promising way to measure the properties of black holes and test General Relativity and alternative theories in the strong-field regime \citep{eht19f,eht22f,Younsi+2023}.

In addition to the astrometry and imaging experiments, the discovery of a pulsar in orbit around Sgr~A* would 
be of particular importance in furthering our knowledge of the black hole and to perform additional tests of gravity theories at an unprecedented level. The extremely high timing precision and stability that can be achieved with radio pulsars has led to very high precision tests of gravity theories (e.g., see \citealt{Kramer+2021,FreireWex2024}).
The information that could be derived via pulsar timing is highly complementary to those from astrometry and imaging experiments.  \cite{Psaltis+2016}, for example, shows how the measurements of the spin and quadrupole moment of Sgr A*, possible with pulsars, have correlated uncertainties that are orthogonal to those of the black hole shadow images.

Over the decades, there has been significant interest in a potentially large but undetected pulsar population towards the GC \citep{Johnston1994,Johnston+1995,CordesLazio1997,PfahlLoeb2004,Wharton+2012,DexterOleary2014}. There is good reason to suspect the presence of a large number of pulsars in the region. Optical photometry of stars close to the GC shows evidence of recent star formation in the past few 100 Myr \citep{Schodel+2020}. This recent star formation could have produced a few hundreds of young pulsars active in the radio band within 1 parsec of the GC. 

The stellar densities in the central parsecs are so high that dynamical encounters between stars can occur at a rate similar to or even higher than that of globular clusters. These types of interactions are typically linked with an overabundance of X-ray binaries \citep{Clark1975} which results in an overabundance of millisecond pulsars (MSP). X-ray binaries do appear to be more prevalent in the central few parsecs from Sgr A* \citep{Munno+2005, Hailey2018}. This suggests that a large number of MSPs may be found in this region. Estimates of this number can be as large as a few thousands \citep{Wharton+2012,Abbate+2018}.

Despite these expectations, dedicated pulsar surveys done in the last 10 years with the most sensitive telescopes available have failed to discover new pulsars in the GC \citep{lde+21,Torne+2021,eatough+2021,Suresh2022,Torne+2023,Desvignes2025,Perez_2026}. The only discovery in the region was possible thanks to the reprocessing of archival observations of the region \citep{Wongphechauxsorn+2024}. This suggests that, in order to find these pulsars, a big leap in sensitivity is needed. In the foreseeable future, the only telescope capable of this leap and that is located in the Southern hemisphere is the SKA-MID.
In Section \ref{sec:science} we report the science results that could be achieved with the discovery of pulsars in orbit around Sgr A* and in the wider GC region. In Section \ref{sec:known_psr} we describe the properties of the pulsars that have already been discovered in this region. In Section \ref{sec:SKA_prospects} we describe the sensitivity limits achievable with the SKA-MID telescope assuming the properties of the staged deliveries AA* and AA4 \citep{braun2019anticipatedperformancesquarekilometre}.

\section{Using Pulsars at the Galaxy's centre} \label{sec:science}

\subsection{Proposed gravity tests}\label{sec:gravity_tests}

%(Contributors: {\bf Zexin Hu}, Lijing Shao)

As excellent clocks in nature, pulsars in the GC or even orbiting Sgr~A* will provide powerful probes of the astrophysical environment and spacetime around the central SMBH, and are a unique test bed for gravity theories~\citep{Paczynski1979,Wex:1995hq,Pfahl:2003tf,Kramer:2004hd,Liu:2011ae,Psaltis:2015uza,Zhang:2017qbb,Hu:2023ubk}. Various studies had shown that even with relatively large timing noise compared to the typical timing precision in binary pulsar systems, pulsars orbiting Sgr~A* can make gravity tests that are unachievable for normal binary pulsars~\citep{Liu:2011ae,Psaltis+2016,Hu:2024blq}. 

Recording and building timing models for the pulse times of arrival (TOAs) from an orbiting pulsar can provide a measurement of the orbital dynamic of the pulsar. At the leading order, the pulsar's motion is described by the standard Keplerian parameters. However, for many systems, the timing precision is high enough to measure the relativistic effects that can be parameterized by the so-called post-Keplerian (PK) parameters~\citep{Damour:1986}. Measurement of $n$ PK parameters can provide $n-2$ independent tests of the gravity theory through the mass-mass diagram. The state-of-the-art illustration of this method of testing gravity theories is from the double pulsar system, PSR~J0737$-$3039A/B, which validates the predictions of GR at better than the $10^{-4}$ level and starts to observe even next-to-leading-order effects~\citep{Kramer:2021jcw}. However, the double pulsar system still represents a relatively weak field case. Consider the depth of the gravitational potential to be $GM/ac^2$, for a total system mass~$M$ and semi-major axis~$a$, with $G$ and~$c$ being the gravitational constants and the speed of light. The current population of binary pulsars probes only to $GM/ac^2 \lesssim 10^{-5.5}$. By contrast, a pulsar with an orbit comparable to the star S0-2 (with an orbital period, $P_b\approx 15$~yr) would probe to $GM/ac^2 \sim 10^{-4.5}$; pulsars with shorter orbital periods could probe to even larger values of $GM/ac^2$.

Pulsars in close orbits around Sgr~A* are extremely suitable for observing several relativistic effects due to the large magnitudes of these effects in a pulsar-SMBH system. The R\"{o}mer delay for a pulsar with 1 year edge-on orbit is of the order of $10^5\, {\rm s}$, which is much larger than the expected timing precision~\citep{Liu:2011ae}. The magnitudes of higher-order post-Newtonian (PN) effects like the first or second order Shapiro delay or Einstein delay \citep{PulsarHandbook} then can be estimated by multiplying the orbital velocity parameter of the system, $\beta_O\equiv(2\pi GM/c^3P_b)^{1/3}\sim 0.02$~\citep{Damour:1991rd}. Eccentric orbits may further enhance the amplitude for these effects. For example, the leading-order Einstein delay can be of order of $20$ minutes~\citep{Liu:2011ae}. Consequently, one expects to measure these relativistic effects to a high precision in a relatively short observation time span like $5\,{\rm yr}$ for a pulsar with $P_b\lesssim 1\,{\rm yr}$. Orbital effects like Lense-Thirring drag and orbital deformation corresponding to the spin-orbital coupling and quadrupole interaction then enable the measurement of the spin and quadrupole moment of Sgr~A* with a fractional precision of $10^{-2}$ to $10^{-3}$~\citep{Liu:2011ae,Psaltis:2015uza,Hu:2023ubk}, and allow for the test of the cosmic censorship conjecture and the no-hair theorem in GR. On the other hand, due to the extreme mass ratio of this system, which is smaller than $10^{-6}$, the orbital decay caused by gravitational radiation only becomes important for very compact orbits ($P_b\lesssim 10\,{\rm hr}$), that are unlikely to be found in the future. It's important to note that the timing models currently used based on the PN formalism will start to show inaccuracies in the case of a bright MSP in close orbit around SgrA* and a more accurate approach based on full GR will be needed \citep{Carleo2025}.

% new text
The values of the spin and quadrupole moment of Sgr~A* that can be measured using pulsars can be compared to the constraints obtained from the EHT images and the orbit of the S2 stars \citep{Psaltis+2016,Ayzenberg2025}. Additional constraints can come from observations with the future Laser Interferometer Space Antenna (LISA, \citealt{Amaro-Seoane+2023}). This instrument has the potential to observe extreme mass ratio inspirals (EMRIs) around Sgr~A* that could provide tight estimates of its spin with a fractional precision of 10$^{-2}$, similar or even better than what is achievable with pulsars \citep{Tahura2022}. The constraints from the different probes are highly complementary and will allow for invaluable tests of self consistency of the models used and increase the confidence in the detection and in the tests of the no-hair theorem.

At present, the prospect of measuring the properties of Sgr~A* with high precision via pulsar timing is based on the assumption that the system is sufficiently unperturbed from the GC environment. However, stars, Dark Matter, and magnetized plasma in fact can cause many perturbations that can be detected in pulsar TOAs and may spoil the measurement of the SMBH properties~\citep{Merritt:2009ex}. Even for a pulsar in a close orbit, if the orbital eccentricity is large, which is favored for measuring the spin of Sgr~A*, the pulsar's orbital motion can be perturbed when it is around the apoastron. A simple strategy to avoid modeling the environmental effects is to use only the timing data corresponding to the periastron passage, where the orbital dynamic of the pulsar is dominated by the SMBH~\citep{Psaltis:2015uza}. However, the spin measurement requires the observation of second-order time derivatives of the orbital parameters in order to break the leading-order degeneracy~\citep{Liu:2011ae,Zhang:2017qbb}, and the measurement becomes much less accurate for the mentioned strategy~\citep{Psaltis:2015uza}. A recent study shows that by combining the timing of two or more pulsars, one can efficiently break the degeneracy with only measuring the leading-order time derivatives and only requires pulsars in moderate orbital periods ($P_b\sim2-5\,{\rm yr}$) thus is compatible with the strategy~\citep{Hu:2024blq}.

By considering the environmental effects or modified gravity effects directly in the timing model, it has been shown that pulsars around Sgr~A* are also unique probe for studying the astrophysical environment around Sgr~A* and testing modified gravity theories \citep{Carleo2023}. Similar to the measurement from tracking S-star orbit~\citep{Heissel:2021pcw}, timing a pulsar orbiting Sgr~A* can constraint the mass distribution around Sgr~A*, which may reflect the dark matter properties~\citep{Hu:2023ubk} or the smooth distributed stars~\citep{Weinberg:2004nj}. Also, significant contribution from another compact object with large mass, for example, an intermediate-mass black hole (IMBH), will leave characteristic signatures in the timing residuals, and their existence can be constrained with pulsar timing~\citep{GRAVITY:2023met}. It also has been shown that pulsar-SMBH system can provide a unique constraint in parameter space for Yukawa gravity, which is a phenomenological model that has a massive graviton~\citep{Dong:2022zvh}. Other modified gravity theories, like bumblebee gravity model, can also be constrained with such systems~\citep{DellaMonica:2023ydm,Hu:2023vsg}.

\subsection{GC interstellar medium studies} \label{ssec:GC_ISM}

An important reason why only a few pulsars have been found in the GC is the scattering caused by the interstellar medium (ISM). Due to multi-path propagation through the magneto-ionized ISM, a portion of the radio signal travels along longer paths, causing asymmetric broadening of the pulse profile. The characteristic timescale of this broadening is known as the scattering time, \(\tau_{\rm sc}\). This is commonly modeled as \(\tau_{\rm sc} = \tau_{\rm sc,1\,GHz} \left(\frac{f}{1\,\mathrm{GHz}}\right)^{-\alpha_{\rm sc}}\), where \(\alpha_{\rm sc} \approx 4\) for a thin-screen approximation. 
%The predictions for the scattering times at 1 GHz can go as high as \(\sim2000\,\mathrm{s}\), effectively smearing out the pulsations of even the slowest pulsars \citep{CordesLazio1997}.
Adopting the GC distance and angular broadening size of Sgr~A* \citep{bdd+14}, and a distance of 130\,pc of the scattering screen from the GC \citep{lazio98}, the scattering time is \(\tau_{\mathrm{sc}} \sim 210\,\mathrm{s}\) at 1\,GHz \citep{mk15}. In this case, finding a MSP in the GC is only possible with observations at a frequency higher than 10\,GHz.

Fortunately, the measurement of the scattering from PSR~J1745$-$2900, at a projected distance of $\sim 0.1$ pc from Sgr~A*, is significantly smaller, \(\tau_{\mathrm{sc}} \sim 1.3\,\mathrm{s}\) at 1\,GHz \citep{Spitler+2014}. 
%On the other hand, measurements of the angular broadening of Sgr~A* and adopting a distance of 130\,pc of the scattering screen from the GC \citep{lazio98} the scattering time is \(\tau_{\mathrm{sc}} \sim 210\,\mathrm{s}\) at 1\,GHz \citep{Bower+14}. 
This discrepancy shows that the scattering scenario around Sgr~A* can be complex. Additional pulsar discoveries close to Sgr~A* will help us create better scattering models and improve our knowledge of the properties of the ISM in the region.

In addition to the scattering properties, eventual discovery of pulsars inhabiting the central parsec will help to determine the gas density and magnetic field strength \citep{Eatough+2015}. The central parsec is characterized by the infall of gas along streams called "mini-spiral" due to their shape resembling spiral arms \citep{LoClaussen1983}. These gas streams are tied with the complex magnetic field in the region \citep{Hsieh2018}. If pulsars are found within these structures, the measurements of Dispersion Measure (DM) and Rotation Measure (RM) can constrain the gas density and the magnetic field, similar to how it was done with pulsars in the nearby Radio Arc non-thermal filament \citep{Abbate+2023}. The time variability and the scattering properties will also be useful to probe the turbulence in the gas.

\subsection{The wider GC and dark matter}

The central few parsecs around Sgr~A* are dominated by the Nuclear Star Cluster (NSC) \citep{Genzel+2010,Neumayer+2020}. This structure has an effective radius of $\sim 4$ pc and a mass of $\sim 2.5 \times 10^{7}$ M$_{\odot}$ \citep{Schodel+2020}. This cluster is mostly made up of old stellar populations, formed more than 10 Gyr ago, but also contains an important fraction of recently formed young stars \citep{Schodel+2020}. A population of young stars only a few Myr old has been found within 1 pc of Sgr A* \citep{Najarro+1997}. The region between 0.04 and 0.2\,pc is populated by a disk of young stars including some born only a few Myr ago \citep{Yelda+2014}. The S-cluster is located even closer to Sgr~A*, at a distance less than 0.04\,pc, and is mostly made up of young and massive stars \citep{EckartGenzel1996,Genzel+2010}. Therefore, the very central regions are the ones where we expect an important fraction of the young pulsars we plan on discovering. A lack of discoveries would have important consequences on the models of stellar evolution.

Similar conclusions can be drawn for the MSPs with a few differences. In order to form MSPs, the neutron star has to spend some time in a binary system where it can accrete material from the companion star and reach the rotational periods of a few ms. Observationally, X-ray binaries containing neutron stars have been seen down to a few parsecs  \citep{Mori+2021} but the angular resolution of X-ray telescopes prevent us from resolving single binaries within the central 0.04 pc. Simulations suggest that the binary fraction of stars can be a few percent outside 0.1 pc but has to decrease inside \citep{Hopman2009,Stephan2019,Panamarev+2019}. This decrease is caused mainly by the disruption of the binary due to the interaction with field stars and the tidal effects generated by the supermassive black hole \citep{Hopman2009}. This suggests that the formation of MSPs might be hampered by the dynamical environment. 

The formation in situ of MSPs is not the only way to place them within the NSC. An alternative formation scenario suggests that MSPs may have formed within globular clusters that collapsed into the NSC \citep{Tremaine+1975,Capuzzo-Dolcetta+1993,Tsatsi+2017,Abbate+2018}. These MSPs, however, are more likely to be deposited in the outskirts of the NSC with only a small fraction of them being able to reach the inner parsec. The effect of mass-segregation is likely to be small since MSPs are not significantly more massive than the rest of the stellar population \citep{Schodel+2020}. The location of where the MSPs are found can therefore give us hints on their potential origin and on the history and dynamical environment of the NSC.

%Non Thermal Filaments
A distinctive feature of the Galactic Centre that is clearly visible in radio images \citep{Heywood+2022}) are the Non Thermal Filaments, thin filaments that run mostly perpendicular to the Galactic disk \citep{Yusef-Zadeh+1984}. An important question still to be answered surrounding these filaments is their origin \citep{Schodel+2024}. One model considers the possibility that the energy source for the filaments is a fast moving compact source like a pulsar \citep{Yusef-ZadehWardle2019}. This model was
%confirmed 
promoted in one recent case where an MSP is seen next to a filament \citep{Yusef-Zadeh+2024,Lower+2024}; although the precise pulsar position still needs to be measured. This suggests that other Non Thermal Filaments could host nearby pulsars making them an important target for our search. 

With a few discoveries of pulsars in the central few parsecs of the GC, we can probe the gravitational potential well in the region. The measured spin-down of these pulsars will have some contributions from the intrinsic spin-down and some from the acceleration imparted on the pulsar by the gravitational potential well of the GC \citep{Perera+2019}. This technique can lead to an independent determination of the total mass of the NSC which can be used to constrain the density and distribution of Dark Matter. These results would complement the on-going efforts of determining the mass distribution in the GC through optical astrometry \citep{Grav:mass_distr2022,Grav:mass_distr2024} using long-baseline optical interferometers like the VLTI (Very Large Telescope Interferometer) or the proposed LPI (La Palma Interferometer).

%({\bf Zexin Hu}, Lijing Shao) 
The presence and nature of Dark Matter can also be probed directly by binary pulsars in the GC region or pulsars orbiting around Sgr~A*. Pulsar with a white dwarf companion can be used to test the universality of free fall towards Dark Matter, which is related to the couplings of Dark Matter to the Standard Model~\citep{Shao:2018klg}. Current timing of binary pulsar system PSR~J1713$+$0747 already provide important improvement over other limits, while binary pulsar systems in the GC region can provide much better constraints due to the large Dark Matter density and acceleration here. As discussed in the previous section, pulsars orbiting Sgr~A* can be a powerful probe of the mass distribution around the SMBH. It has been shown that for cold Dark Matter, the spike structure around SMBH~\citep{Gondolo:1999ef} may provide enough mass of Dark Matter to be detected by timing a pulsar around Sgr~A*~\citep{Hu:2023ubk}. This may help answer the question of whether Dark Matter has a spike or core structure in the GC, the latter one being usually predicted by self-interaction or ultra-light Dark Matter.

%\par ({\bf J. Wongphechauxsorn}, G. Saowanit) 
Moreover, an alternative candidate for Dark Matter is axions; hypothetical, extremely light and weakly interacting particles proposed to solve the strong CP problem in quantum chromodynamics \citep{Peccei:1977hh, Weinberg:1978ma, Wilczek:1978pj}. When axions experience a significant magnetic field, they can undergo conversion into photons \citep{Sikivie:1983ip, Raffelt:2006rj}. The probability of this conversion increases with the  magnetic field strength and dark matter density. Neutron stars are one of the most magnetized observable objects, with a magnetic field of up to $10^{14}$ G in the case of magnetars. Moreover, as mentioned before, the density of Dark Matter is theoretically and observationally high in the GC \citep[e.g.,][]{Hook:2018iia, Millar:2022vhe,Perera+2019}, making the pulsars in the GC an ideal tool for studying the axion-Dark Matter hypothesis. Previously, there have been a few studies searching for emission lines in the spectrum of the magnetar PSR~J1745$-$2900 at various frequencies \citep[see e.g.,][]{Foster_et_al_2020,Darling+2020}. However, due to the limited sensitivity, and the variability of magnetars, discovering more pulsars in the GC can help put a better constraint on this hypothesis. 

% line below not necessary? 
%Therefore, with the existence and future discovery of pulsars in the GC, we can search for the correlated excess emission in their spectra. 

Finally, the presence of Dark Matter can also have an effect on the number and properties of the pulsars found in the GC. For example, the accretion of Dark Matter onto neutron stars, in the case of self-annihilating WIMPs (weakly interacting massive particles), can eventually lead to the collapse into a quark star \citep{Perez-Garcia2010,Herrero2019}. This mechanism would be enhanced towards to the GC where Dark Matter reaches the highest density and can be invoked as an alternative explanation for the low number of observed pulsars. This scenario would open the possibility of observing quark star pulsars that are likely to have a higher maximum spinning frequency and different glitching behaviour compared to neutron stars\citep{Perez-Garcia2010}.

\subsection{Synergies with multi-wavelength and multi-messenger observations}

Exploring the number and distribution of the neutron stars in the GC can help to explain the detection of a $\gamma$-ray excess in the central parts of the Galaxy \citep{Hooper+2011,Hooper+2011b}. The origin of this excess has spiked the interest of many researchers leaving two competing explanations: the annihilation of Dark Matter particles \citep{Hooper+2011, Daylan2016} and the emission from a large population of MSPs \citep{Abazajian2011,Bartels2016}. After more than a decade, this debate is not yet solved \citep{Muru2025} with some studies claiming that the solution involves a combination of both Dark Matter and MSPs \citep{Cholis2022}. The searches with the SKA can provide important limits on the number of MSPs present in the GC and help solve the debate \citep{Calore+2016}.

The presence of a pulsar in extremely close orbit around Sgr~A* would also emit a significant amount of gravitational waves that could be looked at with the upcoming LISA. The emission can either be continuous or burst-like depending on the eccentricity and could, in rare cases, be loud enough to trigger a detection \citep{Kimpson2020}. A detection of the same system in both radio waves and gravitational waves would lead to a wide array of additional tests of gravity \citep{Kimpson2020}.
This synergy would work both ways: a) if a pulsar is discovered in searches with the SKA, the properties derived from pulsar timing would inform LISA of the expected shape of the waveform allowing for the detection 
%also for
of signals that do not reach the signal-to-noise ratio necessary for a blind detection; b) conversely, if a source is found in gravitational wave searches, the SKA dataset can be searched focusing on the specific parameters of the source increasing the chances of detectability. A similar situation could happen if a pulsar is found in orbit around an IMBH in the GC. In this case the gravitational wave counterpart could also be observed with future ground-based gravitational wave observatories like the Einstein Telescope \citep{ETScience2025} or the Cosmic Explorer \citep{CosmicExplorer2021}.

\section{Currently known GC pulsars} \label{sec:known_psr}

\begin{table*}
	\centering
	\caption{Currently known pulsars within a projected distance of 100\,pc from Sgr~A* \citep{Johnston2006,Deneva+2009,Eatough+2013, schnitzeler2016,Abbate+2023,Wongphechauxsorn+2024} We report the rotational period $P$, DM, RM, projected offset from Sgr~A*, the scattering timescale at the frequency reported in parenthesis and notable remarks. The RM is highly variable, the reported value corresponds to the first published measurement and its error in parenthesis. PSR J1745$-$2910 has only been detected in the discovery observations with GBT \citep{Deneva+2009} and its properties and position are not well known.}
	\label{tab:known_pulsars}
	\begin{tabular}{lcrrcrc} % four columns, alignment for each
		\hline
        \multicolumn{1}{c}{Pulsar Name} & \multicolumn{1}{c}{$P$}  & \multicolumn{1}{c}{DM} & \multicolumn{1}{c}{RM}& \multicolumn{1}{c}{Sgr~A* offset} & \multicolumn{1}{c}{$\tau_{\rm sc}$} & \multicolumn{1}{c}{Remarks} \\     
         \multicolumn{1}{c}{PSR}& \multicolumn{1}{c}{(s)} & \multicolumn{1}{c}{(pc cm$^{-3}$)} & \multicolumn{1}{c}{(rad m$^{-2}$)} & \multicolumn{1}{c}{(arcmin / pc)} & \multicolumn{1}{c}{(ms)} &  \\     
        \hline
        J1746–2849 &  1.48 & 1360 &  $10,104 (100)$ & 12.3 / 29 & 266 (1.5\,GHz) & - \\
        J1746–2850 &  1.08 &  941 & $-12,363 (40)$ & 11.3 / 27 & - & magnetar-like \\
        J1746–2856 &  0.95 & 1155 &  $13,253 (50)$ &  15.6 / 37 & 450 (1\,GHz) & - \\
        J1745–2900 &  3.76 & 1778 & $-66,960 (50)$ &  0.04 /  0.1  & 1300 (1\,GHz) & magnetar \\
        J1745–2912 &  0.19 & 1106 &     $-535 (100)$ & 14.5 /  34 & 750 (1\,GHz) & - \\                       
        J1745–2910 &  0.98 & 1088 &  -  & 12.9 /  31 & - & - \\
        J1746–2829 &  1.89 & 1309 &     $-743 (14)$ & 31.8 / 76.0 & 67 (1 GHz)& magnetar-like\\
		\hline
	\end{tabular}
\end{table*}

The lack of pulsar discoveries in the GC in the first decades of observations raised particular interest in the community \citep{Johnston1994, Johnston+1995,CordesLazio1997}. The large distance from Earth coupled with the expected high DM and strong scattering made it particularly difficult to discover any pulsars in the region. The first two pulsar discoveries \citep{Johnston2006} only happened from observations at the Parkes telescope at higher frequencies, 3100\,MHz, where the effect of the DM and scattering are reduced. After these, three more discoveries were made thanks to observations at GBT also at a marginally higher frequency of 2000\,MHz \citep{Deneva+2009}. An X-ray outburst close to Sgr~A* lead to the discovery of a magnetar that was later found to be active also in radio \citep{Kennea2013,Mori+2013,Eatough+2013}. Finally, the last discovery in the region happened in a reprocessing of Parkes data using the Fast Folding Algorithm (FFA) \citep{Wongphechauxsorn+2024}. Currently, therefore, there are 7 known pulsars located within a projected distance of 100\,pc from Sgr~A*. These pulsars have among the highest values of DM and RM ever measured for pulsars \citep{schnitzeler2016}. Remarkably, apart for the known magnetar PSR~J1745$-$2900, two more pulsars, PSRs~J1746$-$2850 and J1746$-$2829, show some properties that are similar to those of magnetars \citep{Deneva+2009, dexter+2017, Wongphechauxsorn+2024}, like long rotational periods, flat radio spectra ($\alpha<-1.0$), strong fluctuations in flux and high magnetic fields. This might suggest that the formation of magnetar-like neutron stars is more common in the GC than it is in the rest of the Galaxy \citep{DexterOleary2014} and could allow for tests of the formation scenarios of magnetars.
Some of the formation scenarios that have been proposed range from highly magnetised progenitor stars \citep{Hu2009} to interactions involving tight massive binaries \citep{Fuller2022,Ablimit2022}.  However, a recent study suggests that magnetar formation might be more common throughout the Galaxy than previously thought \citep{Pardo-Araujo2026} and the observed prevalence in the GC could be due to selection effects.

%\cmts{RPE: Do both these pulsars have timing solutions to derive spin down?} \cmts{JW: For J1746-2829, yes we have a partial timing solution for few years. It is not clear if they have that for J1746-2850 but we can see a strong period change.}

Despite the extensive searches, no MSPs have been found within 100\,pc from Sgr~A*. One reason for this can be the strong scattering that affects not only Sgr~A* but the entire GC region. In Table \ref{tab:known_pulsars} we show the values of the scattering timescales of the closest pulsars. Even for PSR~J1746$-$2829, at a projected distance from Sgr~A* of 76\,pc, the scattering timescale at 1\,GHz is $\sim 70$\,ms \citep{Wongphechauxsorn+2024}. The closest known MSP is PSR~J1744$-$2946 \citep{Lower+2024} located at $\sim 120$ pc from Sgr~A*. For this pulsar, the scattering timescale extrapolated down to 1 GHz would be $\sim 40$ ms. This scattering would make any MSP completely undetectable at low frequencies. However, even high-frequency surveys, which highly reduced the impact of scattering, were unable to discover new pulsars \citep{mkf+10, lde+21, Torne+2021, eatough+2021, Suresh2022, Torne+2023}. The conclusion is that the non-detections are mainly a result of an insufficient sensitivity to the pulsar signals -- a consequence of the typical steep spectrum of pulsars and the increased frequencies necessary to overcome the scattering. This is one of the main reasons why the SKA-MID's superb sensitivity is needed for sensitive-enough pulsar surveys in the Galactic Centre \citep{Eatough+2015, mk15}.

\subsection{PSR~J1745$-$2900}
\begin{figure}
    \centering
    \includegraphics[width=0.48\textwidth]{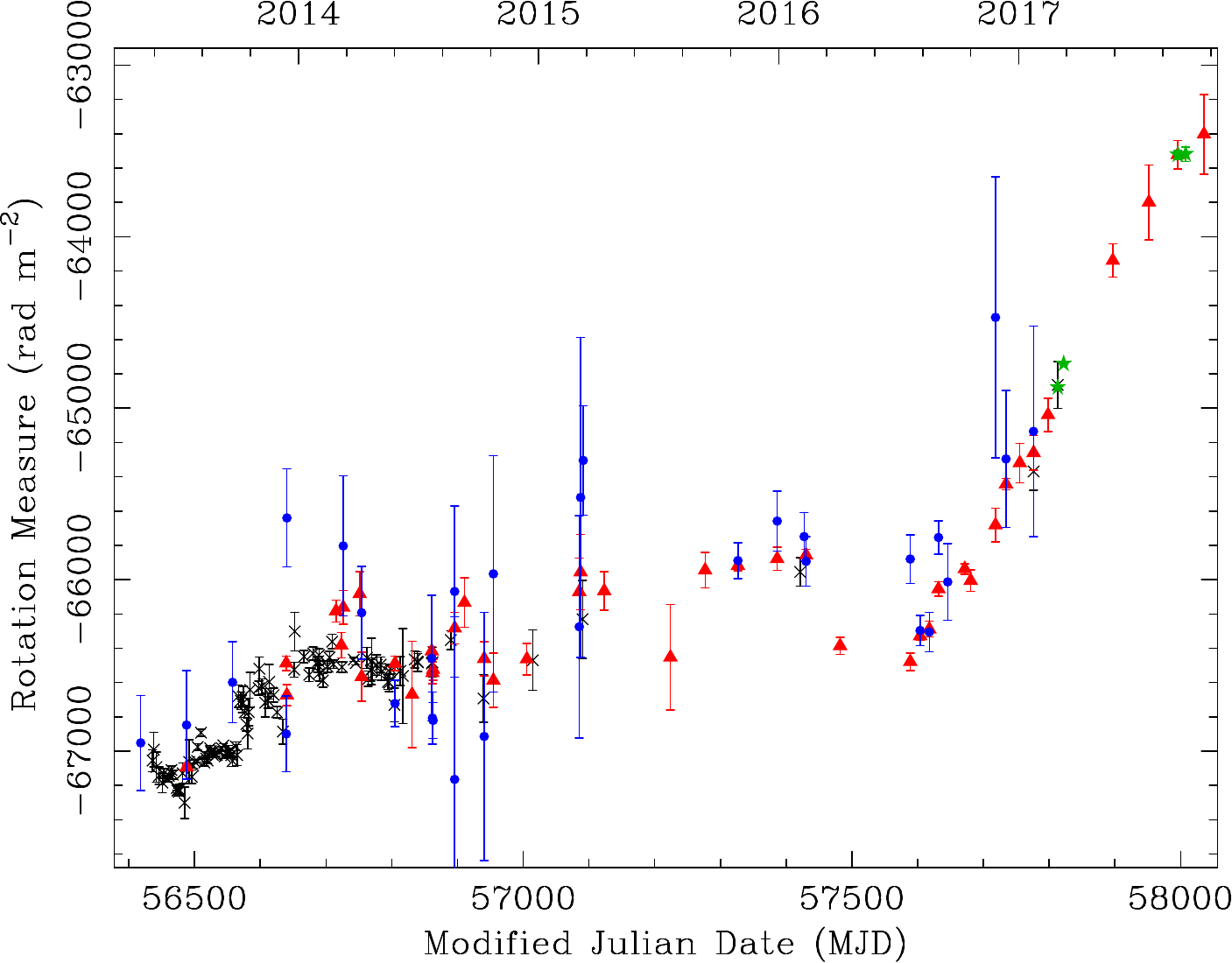}
    \caption{Evolution of the Faraday RM towards the line-of-sight of PSR~J1745$-2$900. Black crosses, red triangles, blue points, and green stars indicate RM values obtained with the Nan\c cay Radio Telescope at  2.5\,GHz and Effelsberg at 4.85, 8.35 and 6\,GHz, respectively. Adapted from \citet{Desvignes+2018}.}
    \label{fig:RM}
\end{figure}

Of the known GC pulsars, PSR~J1745$-$2900 arguably holds extra significance due to its proximity to Sgr~A* of just 0.3 light years in projection \citep{Rea+2013}. Confidence in its location in the GC was boosted by the measurement of the highest known RM after Sgr~A* itself \citep{Eatough+2013}. Since
then, large RM variations of order 5\% were reported in 2018 (Fig.~\ref{fig:RM}), arguing for small-scale ($\gtrsim$ 2 a.u.) magneto-ionic fluctuations local to the GC environment, based on measurements of the pulsars proper motion \citep{Desvignes+2018,Bower+15}. 
%\citep{Desvignes+2018}. Mention further RM changes here?
Such large and variable fluctuations in magneto-ionic properties have only been seen in pulsars orbiting Be stars \citep{Johnston+1996} and extragalactic Fast Radio Burst (FRB) sources, indicating a tantalizing connection to some of the FRB host environments \citep{michilli2018}.    

The initial measurement of pulse scatter broadening in PSR~J1745$-$2900 \citep{Spitler+2014} indicated a value several orders of magnitude below that of previous predictions \citep[e.g.][]{CordesLazio1997,lazio98}. The scattering appears to be caused by a single thin screen of material \citep{Wucknitz2014}. Combination of the pulse scatter broadening measurements with angular scatter broadening obtained  with the Very Long Baseline Array (VLBA) indicate that the screen is unexpectedly distant from the magnetar and GC at 5.9 $\pm$ 0.3 kpc \citep{Bower+14}. This result has been corroborated by a long-term astrometric campaign of the pulsar \citep{Bower+2025}. The angular broadening scale has not shown changes during the almost 500 days of the campaign, in contrast with the large variations of RM. This supports the idea that material responsible for the scattering is much more distant than the material causing the RM changes. The scale over which the scattering changes must be larger than the distance traveled by PSR~J1745$-$2900 during the observing campaign ($\sim 200$ AU).

While it is thought PSR~J1745$-$2900 is still too distant for most gravity tests of Sgr~A* \citep{pen_and_brod_2014}, its detection has already illustrated the potential of pulsars in this region to reveal a large amount of previously unknown properties of the Galaxy's central region.

%\section{Prospects for the staged GC pulsar SKA search} 
\section{Prospects for the staged SKA GC pulsar search} 
\label{sec:SKA_prospects}

%Consider both deep searches around Sgr A* of multiple hours of integration and shallow search with short integration times of the surrounding 50-100 pc.

\subsection{Sensitivity curves}

The sensitivity of a pulsar survey can usually be estimated by using the radiometer equation applied to the observation of pulsars, as in e.g., \citet{mkf+10} and \citet{Eatough+2015}. This involves calculating the effective width of the pulsar signal which is a quadrature sum of the intrinsic pulse width together with some external effects that can broaden the pulse profile, such as DM smearing, temporal scattering, finite sampling interval, etc. For the case of pulsar searching in the GC, the temporal scattering could be so severe that it spreads the power of the pulsar signal over more than one rotational period of the pulsar. Under this circumstance, the ordinary radiometer equation will return an imaginary value for the sensitivity calculation. Therefore, to avoid this issue, here we employ the survey sensitivity formula from \citet{cc97} as below:
\begin{equation}
    S_{\rm min} = \frac{\sigma\alpha\cdot{\rm SEFD}}{\sqrt{2BT_{\rm int}}}\cdot\frac{1}{H},
\end{equation}
where $\sigma$ is the detection threshold for the survey, $\alpha=\sqrt{1-\pi/4}$ is a coefficient related to the RMS noise in the Fourier Transform of the intensity, $B$ is the bandwidth and $T_{\rm int}$ is the total integration time. The factor $H$ is written as:
\begin{equation}
    H=\frac{1}{\sqrt{N_{\rm h}}}\sum_{l=1}^{N_{\rm h}}R_l,
\end{equation}
where $N_{\rm h}$ is the number of harmonics summed in an FFT search and $R_{l}$ is the ratio of the Fourier Transform amplitude for the $l$-th harmonic to the zero frequency amplitude. Assuming a Gaussian profile shape, $H$ can be expressed by
\begin{equation}
    H=\frac{1}{\sqrt{N_{\rm h}}}\sum_{l=1}^{N_{\rm h}}e^{-(\pi\epsilon l/2\sqrt{\rm ln2})^2},
\end{equation}
where $\epsilon\equiv w/P$ is the duty cycle. Once the survey sensitivity is calculated, one can use that to estimate how deeply the survey can probe the GC pulsar population, by deriving the luminosity threshold at a given frequency and compare it with the anticipated luminosities of a large pulsar population at the same frequency. Here, we follow a strategy similar to that in \cite{lde+21}. From the ATNF pulsar catalog \citep{mhth05}, we chose all pulsars with flux density measurements around 1.4\,GHz or above as our samples. This kept approximately 57\% of over 4100 pulsars in total. Then for pulsars with a reported spectral index, we extrapolated the flux density to 10\,GHz assuming the power-law spectrum extends up to that frequency. Otherwise, we employed a random value drawn from a normal distribution (with mean of $-1.6$ and standard deviation of 0.54) obtained by \cite{jvk+18}. The distance used to calculate the luminosity at the distance of GC are mostly based on DM and the YMW16 Galactic free-electron density model \citep{ymw17}. 

%The distance used to calculate the luminosity at the distance of GC are mostly based on DM and the YMW16 Galactic free-electron density model \citep{ymw17}. 

\begin{figure*}
    %\centering
    \hspace*{-1.5cm}
    \includegraphics[scale=0.49]{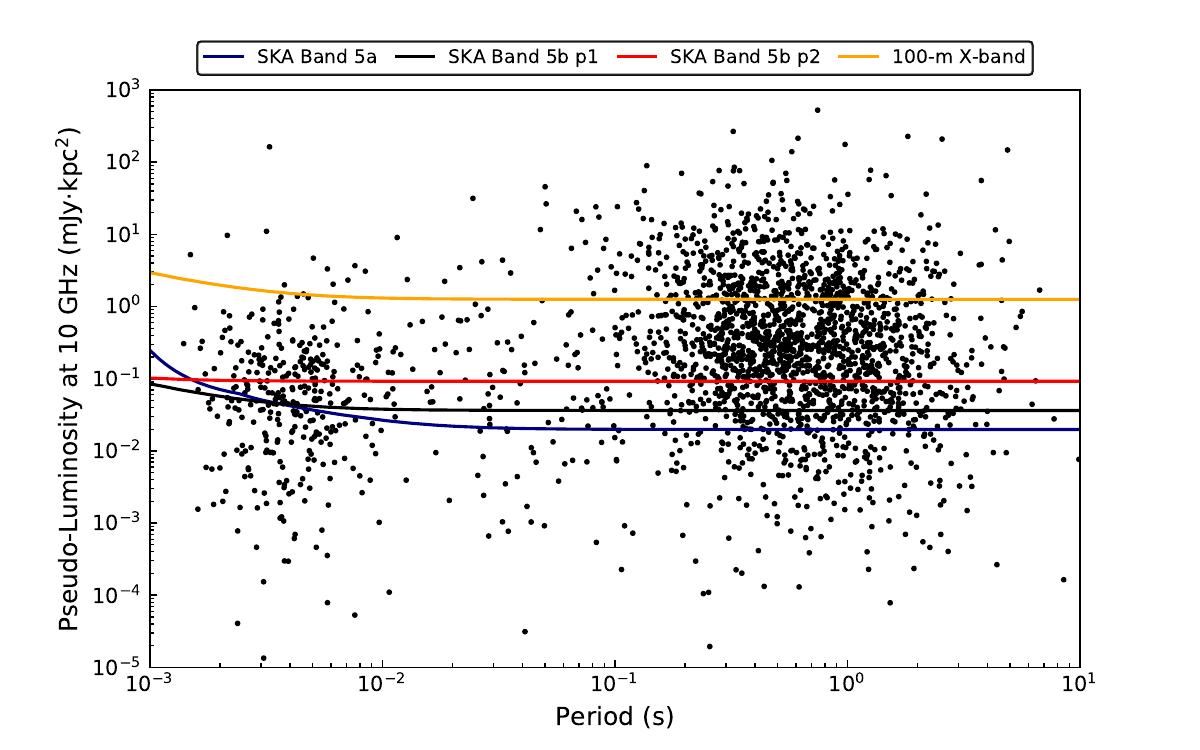}
     \hspace*{-1cm}
    \includegraphics[scale=0.49]{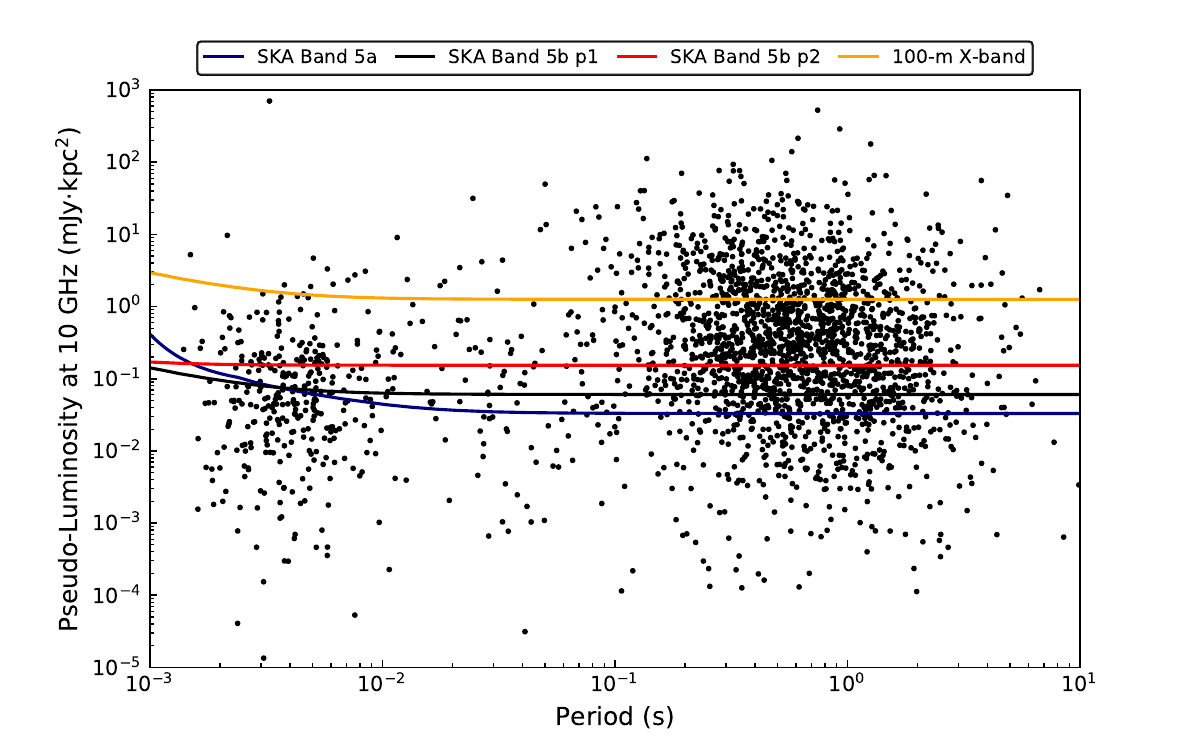}
    \caption{Sensitivity of SKA GC surveys for AA4 (left) and AA* (right) configurations, using the GC magnetar scattering timescale. The integration time was assumed to be 4\,hr. The central frequency of band 5a, band 5b part1, band 5b part2 are 6.6, 9.55 and 14.15\,GHz, respectively, each with a 2.5-GHz bandwidth. This avoids the radio interference within 10.7--12.7\,GHz caused by the Starlink satellites. The sensitivity of AA4 at these frequencies are approximately 1250, 1120, 890\,m$^2$/K, respectively, and AA* possesses 60\% of AA4's sensitivity. For the sensitivity estimate of survey with a 100-m dish, we assumed a system SEFD measured during GC observations with the Effelsberg radio telescope by \citet{eatough+2021}. For the case of AA4, the survey is anticipated to detect 84\%, 79\%, 66\% of the whole GC pulsar population, and 60\%, 59\%, 43\% of the GC MSP population. For AA*, these are 78\%, 71\%, 56\% for the whole population, and 50\%, 46\%, 31\% for the MSP population.}
    \label{fig:weak_scatter}
\end{figure*}

\begin{figure*}
    \hspace*{-1.5cm}
    \includegraphics[scale=0.49]{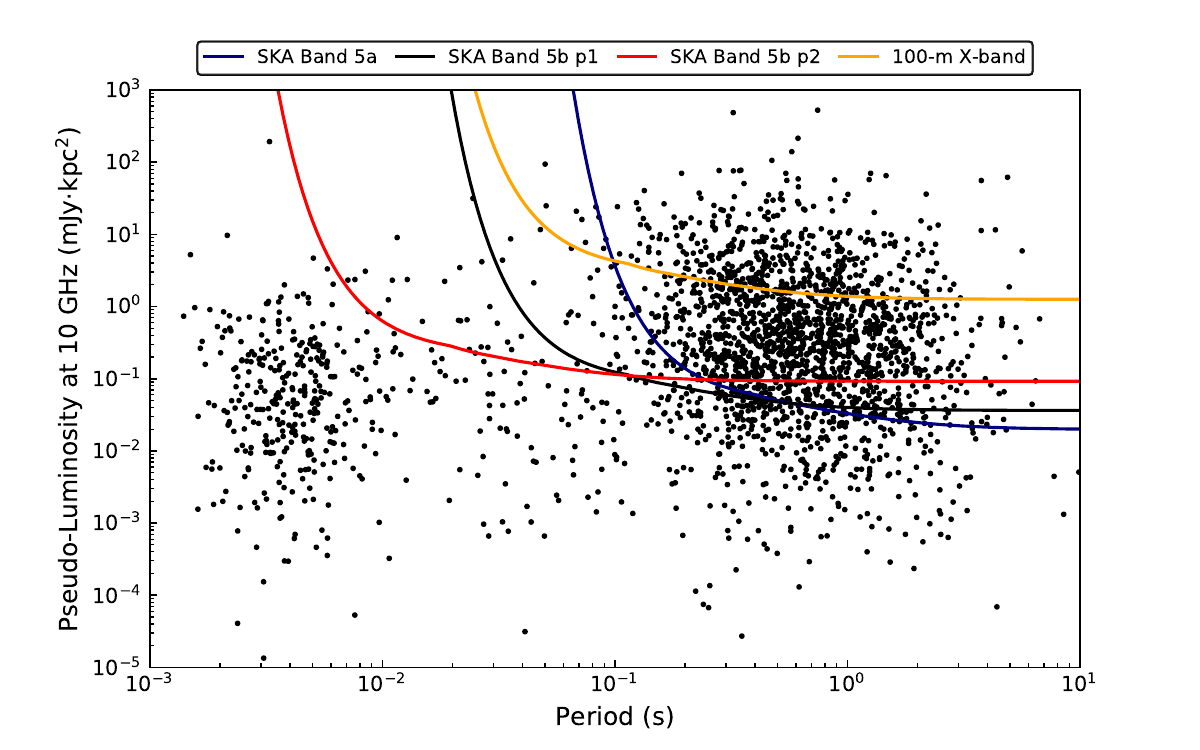}
         \hspace*{-1cm}
    \includegraphics[scale=0.49]{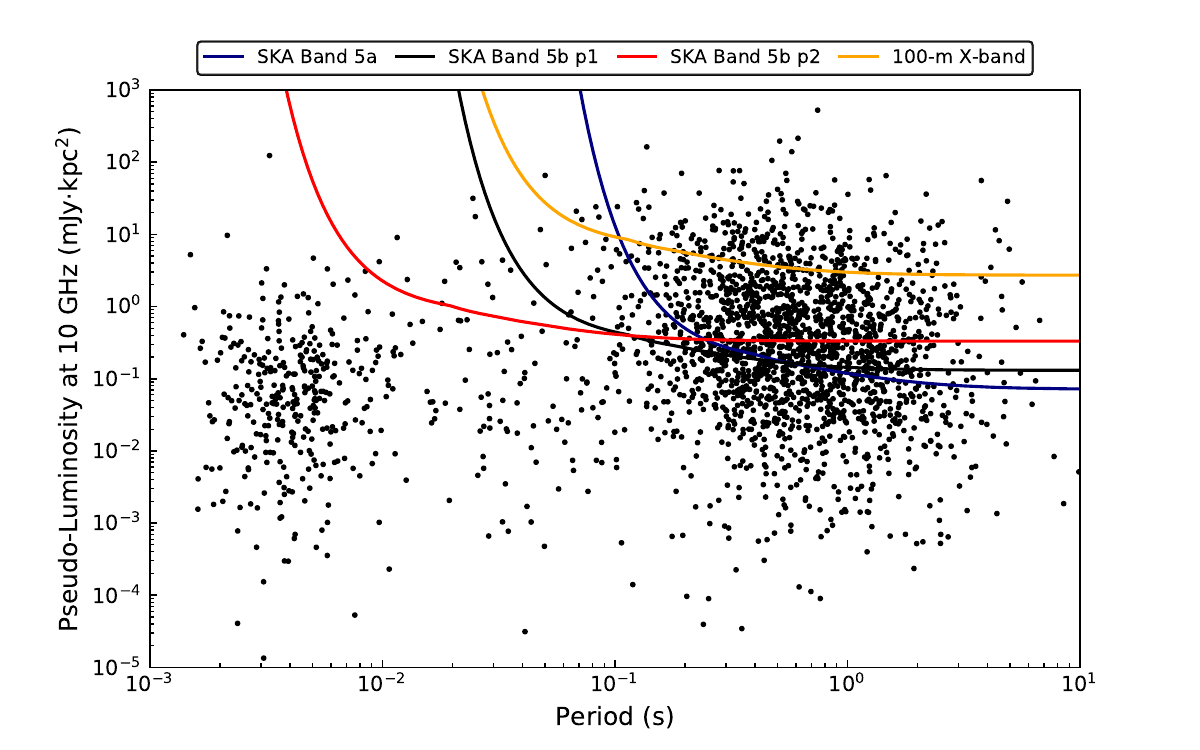}
    \caption{Sensitivity of SKA GC surveys for AA4 (left) and AA* (right) configurations, assuming a strong scattering scenario as mentioned in Section~\ref{ssec:GC_ISM}. The observational setup is the same as in Figure~\ref{fig:weak_scatter}. The survey is anticipated to detect 64\%, 65\%, 60\% of the whole GC pulsar population when using AA4, and 58\%, 60\%, 52\% when using AA*. As for MSPs, only the survey at Band 5b part 2 would be able to detect a very small fraction of MSPs, namely 8\% for AA4 and 6\% for AA*.}
    \label{fig:strong_scatter}
\end{figure*}

Figures~\ref{fig:weak_scatter} and \ref{fig:strong_scatter} show the GC survey sensitivities for a few options in SKA band 5 under both the AA4 and AA* (73\% of AA4) array 
%configuration
stages
\citep{braun2019anticipatedperformancesquarekilometre}, assuming the GC magnetar scattering and a strong scattering scenario, respectively. Three cases are considered: band 5a (5350 - 7850 MHz), the low part of band 5b (8300 - 10800 MHz), called band 5b part 1, and the upper part of band 5b (12900 - 15400 MHz), called band 5b part 2. Under the assumption of GC magnetar scattering, the surveys using AA4 sensitivities will detect up to 84\% of the pulsar population, at the GC distance, and up to 60\% of the MSP population in the GC. The option that gives the highest detection fraction is band 5a. For AA* sensitivities, the detection fractions are generally $\sim$10\% lower, but would still be able to detect half of the MSP population. In contrast, for the case of strong scattering, the detection fraction for the whole GC pulsar population drops from 4 to 20\%, depending on the choice of the central frequency. The surveys would mostly not be sensitive to detect MSPs in the GC, except for Band 5b part 2 which though may only be able to detect a very small fraction of GC MSPs. This result is not surprising, because as already pointed out by \citet{mk15}, the optimal survey frequency for GC MSPs in this case is somewhere above 20\,GHz. 

%\pt{What Tsys are used for the calculations? I think the ones without contribution from GC(?) If we add T$_{GC}$ to Tsys, the lines are about a factor 4-7x higher and changes the results (\% of detectable pulsars) in a non-negligible way.}

%\kl{Our beam is tiny so most Tsys would be resolved out. I can add the flux density of Sgr A* to the SEFD, but that would increase the SEFD by a fraction.}

It should be noted that the sensitivity of a GC pulsar survey in practice is usually more complicated than the theoretical estimation above. For instance, baseline variations due to red-noise in the data is a commonly seen feature and can significantly hinder the FFT search sensitivity to pulsars with a comparatively long rotational period (e.g., $P\gtrsim1$\,s). Instrumental artifact could also exhibit some periodicities and spread additional power across the Fourier spectrum. Meanwhile, for observations with multiple beams, the baseline subtract technique could be applied to significantly lower the red-noise amplitude in the Fourier spectrum. The sensitivity of an FFA search is also largely different from an FFT search. A more reliable sensitivity estimate of a pulsar survey would usually come from injection of fake signals into the real data and employ the search pipeline to recover the signal and calculate its detection significance. This has already been conducted in a number of previous pulsar surveys in the GC, such as \citet{lde+21,Torne+2021,eatough+2021,Torne+2023}. The sensitivity of future SKA surveys can be evaluated within exactly the same framework. 

%\cmts{from KJL: I think the sensitivity curve is wrong. Due to the red noise in searching data, the sensitivity of searching for long period pulsar  will be worse. see eg. Lazarus.,P. et al., 2018, Cordes, J. M., \& Chernoff, D. F. 1997}

%\cmts{from JW: How about we said that this is just an theoritical value, while it is worst towards longer period due to rednoise ? }

\subsection{Search Strategies}
\label{sec:search_strategies}

Detecting pulsars in the GC presents a multifaceted challenge due to both intrinsic and extrinsic factors. Typical radio pulsars exhibit spin periods between \(\sim\)1.5\,ms and a few seconds, with duty cycles of approximately more than 0.1\%. The environment in the GC introduces severe propagation effects and binary-induced timing distortions that must be mitigated to maximize the sensitivity of the searches \citep[see e.g.,][]{eatough+2021}. Furthermore, some pulsars could be in relativistic orbits around Sgr~A*, with orbital periods ranging from hours (in the case of pulsar-black hole binaries near the last stable orbit) to decades. In this section, we outline the search strategies used to mitigate these effects on pulsar detectability.

Detecting pulsations is inherently difficult due to the faint and often noise-like nature of single pulses. Sensitivity can be improved by accumulating signal power through periodicity searches. Most known pulsars have been discovered using the Fast Fourier Transform \citep[FFT,][]{cooleyA+1965} applied to dedispersed time series data, followed by folding at the candidate period. However, since pulsar signals are often Gaussian-like, their power is distributed over a small number of Fourier bins, with a peak at the fundamental frequency. This spread can be mitigated using incoherent harmonic summing, which adds power from higher harmonics back to the fundamental frequency \citep{taylor+1969}. However, this approach is not always sufficient due to many reasons, such as too many harmonics outside the Nyquist frequency range. Most of the GC pulsar search to date are using FFT as a main algorithm for periodicity search \citep[e.g.,][]{Johnston2006,Deneva+2009,eatough+2021,lde+21,Torne+2023}

An alternative way to search for pulsations is the FFA, first proposed by \citet{Staelin1969}. The FFA folds time series at many trial periods directly, producing higher sensitivity to long-period or low-duty-cycle signals. While traditionally more computationally expensive than FFTs, recent algorithmic improvements and enhanced computing resources have enabled efficient implementations suitable for large-scale surveys \citep{Morello2020}. For example, \citet{Morello2020} showed that for a 10\% duty-cycle signal, FFTs recovered only \(\sim80\%\) of the S/N achieved with the FFA and less for smaller duty cycles and narrower pulse profiles. Since the duty cycles are proportional inverses  with the period of the pulsar, the FFA is better for longer-period pulsar searches. So far, only two surveys in the GC have been utilizing the FFA, leaving a large possible undiscovered population of pulsars in the GC \citep{Torne+2023,Wongphechauxsorn+2024}.

Both FFT and FFA are searching for pulsation under the assumption that the period of the pulsations remains constant over the observation. However, the apparent spin period of pulsars can be changed due to the Doppler effect caused by the motion of the pulsar around its potential companion, reducing the detectability of the pulsar. As we are searching for pulsars orbiting around the SMBH, we need to take into account the binary motion. Moreover, some of the pulsars in the GC might have their own companion due to the dynamical environment in the GC. The ideal method would be to search for five Keplerian orbital parameters. Then, the timeseries can be resampled at each data point to align with the rest frame for each set of parameters. When these parameters are optimal, the signal-to-noise ratio (S/N) of the signal will be at its maximum. \citet{balakrishnan+2022} implemented a radio pulsar search using this principle, which is known as a template-bank search. However, this method requires a significant amount of computational resources, which is not sustainable for large-scale pulsar surveys. 

Usually, for binary pulsar searches, the orbital motion is compensated for by searching in the space of pulsar's spin period and its derivatives which correspond to the acceleration of the pulsar induced by its companion. When the time span of the search is significantly less than the orbital period, one can assume that pulsars move with a constant line-of-sight acceleration. With this assumption, the search will now depend solely on one parameter, acceleration $a$. Similar to template bank search, the timeseries will be resampled to each value of $a$. The optimal acceleration will yield the maximum S/N \citep[see e.g.][]{Eatough+2013b}. Alternatively, \citet{andersen+2018} implemented a Fourier domain acceleration and jerk search accounting for the distinctive smearing pattern in the Fourier spectrum. Following the notation in \citet{andersen+2018}, this is denoted by the $z_{\rm max}$ and $w_{\rm max}$ parameters for Fourier domain smearing caused by acceleration and jerk respectively. The acceleration search is less computationally expensive but has a limitation where the observation time is, e.g., 10 times the shortest detectable binary period to satisfy the constant acceleration assumption \citep{ransom+2003,ng+2015}.
%As for higher order search, the rate of acceleration change i.e. `jerk' can be also searched in a similar manner to acceleration search. For example, the Fourier domain jerk search searches for additional smearing patterned of $w$ bins \citep[see][for details]{andersen+2018}.
In \citet{andersen+2018} they demonstrated that this assumption holds under the condition that the observation time is $\sim$ 15 per cent of the orbital period.  

\begin{figure*}
\centering
    \includegraphics[scale=0.4]{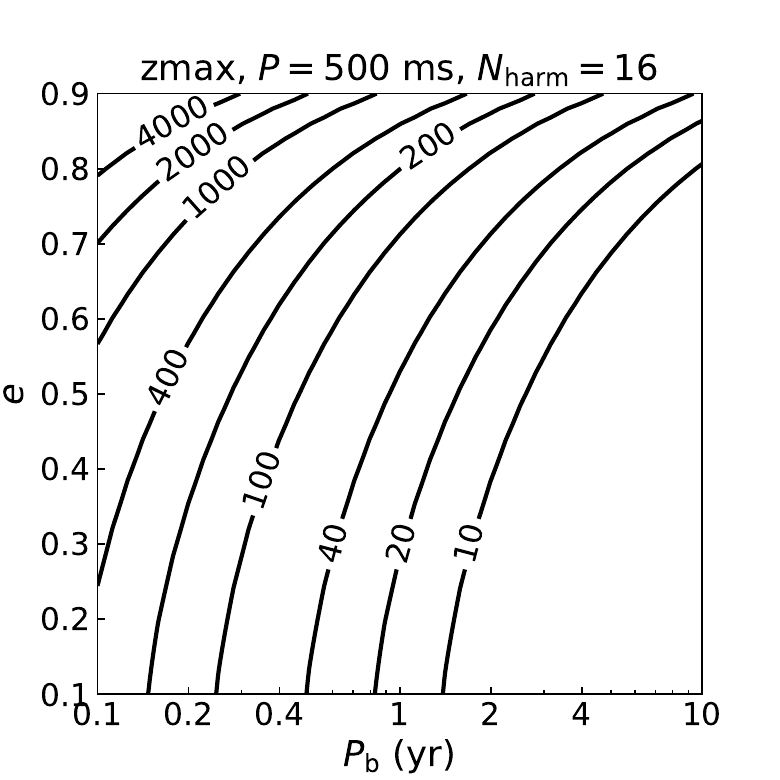}
     \hspace*{-0.4cm}
    \includegraphics[scale=0.4]{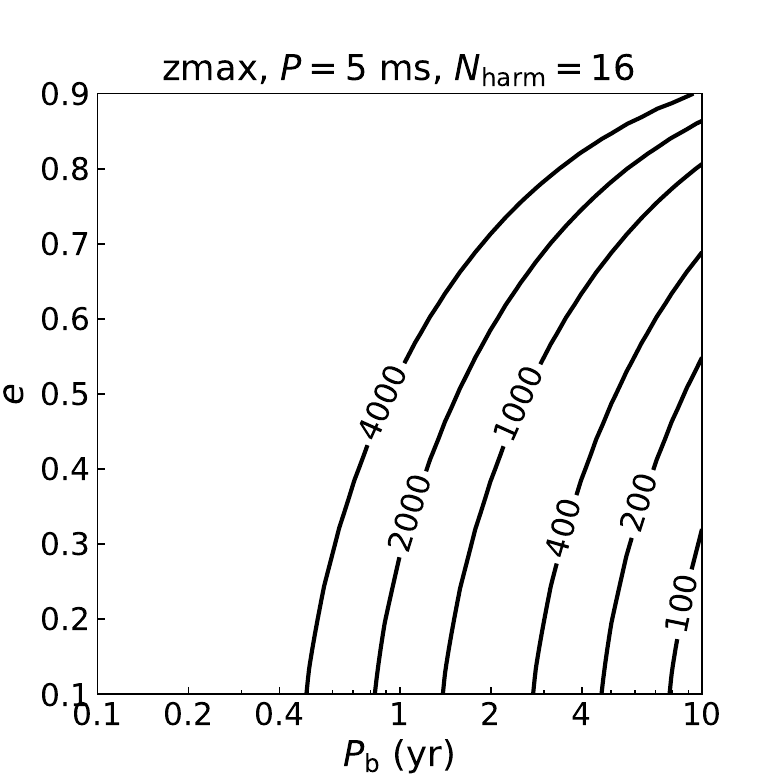}

    \includegraphics[scale=0.4]{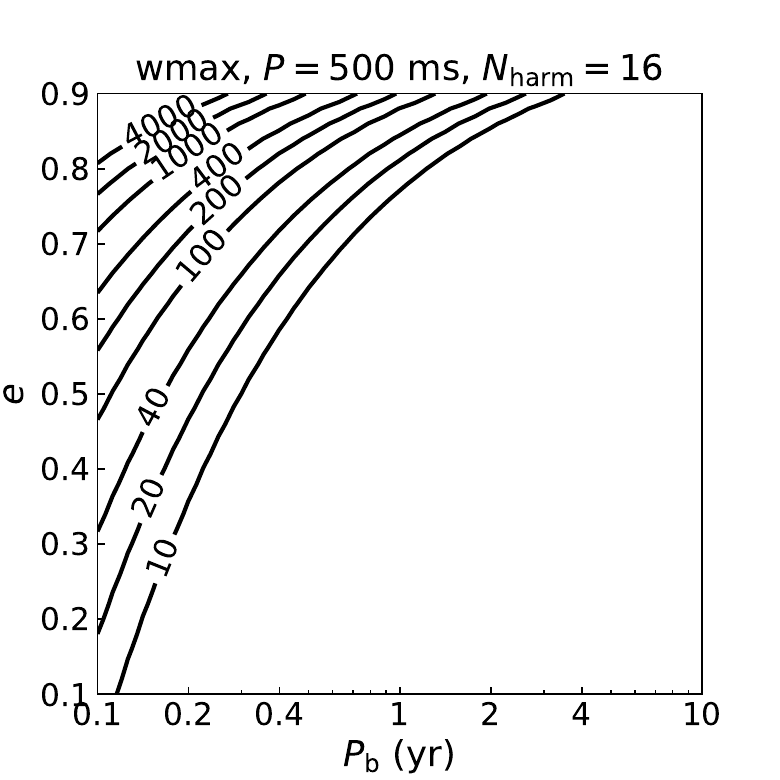}     
    \hspace*{-0.4cm}
    \includegraphics[scale=0.4]{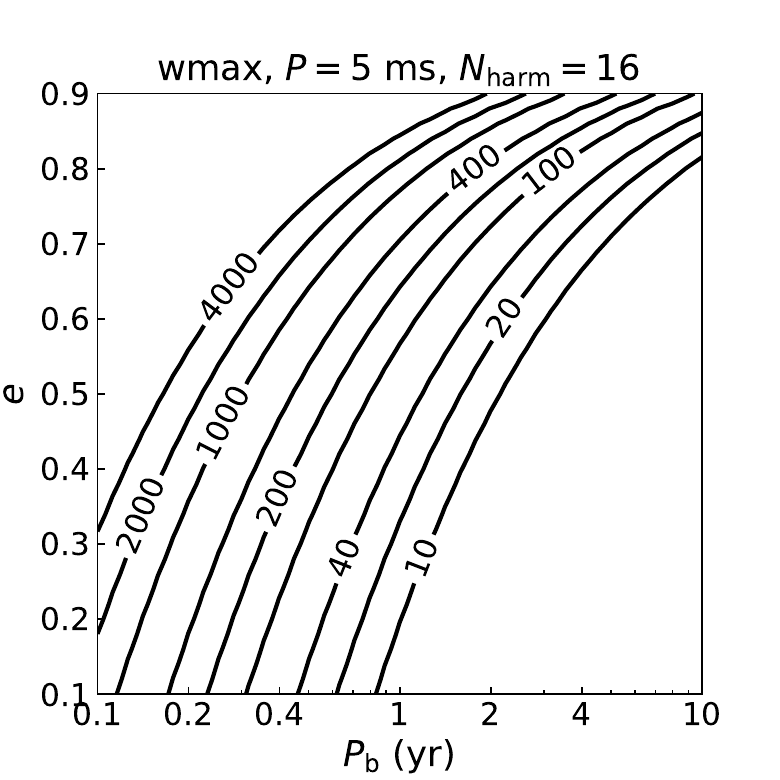}
\caption{$z_{\rm max}$ (upper row) and $w_{\rm max}$ (lower row) values required in \textsc{presto} by a 4-hr search to incorporate the maximum acceleration and jerk for a range of pulsar orbits around the Sgr~A*. The spin period of the pulsar was assumed to be 500 and 5\,ms for case of an ordinary pulsar and an MSP, respectively. The number of harmonics summed in the search was set to be 16. \label{fig:zwmax}}
\end{figure*}

% expected range for SMBH 
%Plot showing high demand for zmax and low demand for wmax. 
As for the case of a pulsar in orbit with the Sgr~A*, this forms a very special ``binary'' pulsar system. The ``companion'' is extremely massive and the pulsar endures significant acceleration when orbiting around Sgr~A*. For instance, in a circular orbit with a period of 1\,yr, the line-of-sight acceleration can already be as large as 1\,m/s$^2$. For an orbit of period of 0.1\,yr, it is then approximately 20\,m/s$^2$. Meanwhile, for the vast majority of possible orbits, the period is beyond the order of 10--100\,day, meaning the duration of each observation is far less than the orbital period. Therefore, for an acceleration search, one would need to apply a significantly larger value of $z_{\rm max}$ rather than $w_{\rm max}$, to compensate for the apparent the orbital motion of the pulsar \citep{eatough+2021}. This is the opposite to the case of a compact binary pulsar system \citep[e.g.,][]{andersen+2018}. Generally, the ratio of the maximum of absolute $z_{\rm max}$ and $w_{\rm max}$ out of an entire orbit, is larger for longer integration time, shorter orbital period and higher orbital eccentricity \citep{lde+21}. Figure~\ref{fig:zwmax} gives the maximum absolute $z_{\rm max}$ and $w_{\rm max}$ values required to cover the maximum possible acceleration and jerk in a range of orbits, for a search with integration time of 4\,hr. It can be seen that for an ordinary pulsar ($P=500$\,ms), $z_{\rm max}$ is already required for an orbital period less than a few years. At an orbital period of 0.4\,yr and a modest eccentricity of 0.5, $z_{\rm max}$ of 100 is needed. On the other hand, $w_{\rm max}$ is only required for orbits with either a short orbital period (e.g., $\lesssim0.2$\,yr) or a large eccentricity (e.g., $\gtrsim0.8$). For an MSP ($P=5$\,ms), a $z_{\rm max}$ of a few thousands is already in need for an orbital period of a few years, where $w_{\rm max}$ just starts to become significant. 

%A paragraph about polarisarion search 
In addition to searches using total intensity data, it is also possible to use the polarization components to conduct the search. As discussed earlier, in time-domain periodicity searches, it is common to have baseline variation in the data, which manifests itself as red noise in the Fourier spectrum. This hinders the practical sensitivity of the data in the search, in particular for slow pulsars \citep[e.g.,][]{lbh+15}. However, as shown in \citet{lde+21}, timeseries of the Stokes components, namely $Q$, $U$ and $V$, are significantly more resistant to red noise in the data. If the pulsar signal is significantly polarized (e.g., PSR~J1745$-$2900), then for polarization components, the periodicity search may return comparative or even higher detection significance. Therefore, periodicity search using polarization component would be a good complement to those conducted using the total intensity timeseries. 

A different way to approach pulsar searches is in the image domain \citep{Dai+2017, Frail+2024}. Because the image domain is less affected by pulse shape distortions caused by scattering, and orbital motion causes no effect, they provide a means to explore regions of previously inaccessible search space. This applies in particular to short-orbital period binary systems and to the fastest MSPs. Pulsar candidates identified in the image domain can then be followed up with time-domain observations to confirm pulsations, allowing resources to be focused on areas with higher likelihoods of discovery. This can be achieved by identifying in the image domain unresolved point sources with steeply negative spectra and are circularly polarised \citep{Wang+2022,Sengar2025}.  

The use of the sensitive SKA-MID, combined with the high frequency band, will offer further advantages as they are less impacted by background contributions to the system temperature compared to single-dish telescopes \citep{Anantharamaiah1989}. Additionally, advanced techniques, such as the ability to subtract the incoherent beam, can significantly reduce red noise and improve detection sensitivity.

\subsubsection{Search around Sgr A*}
\label{subsection:SGRsearch}

\begin{figure*}
    \centering
	\includegraphics[scale=0.6]{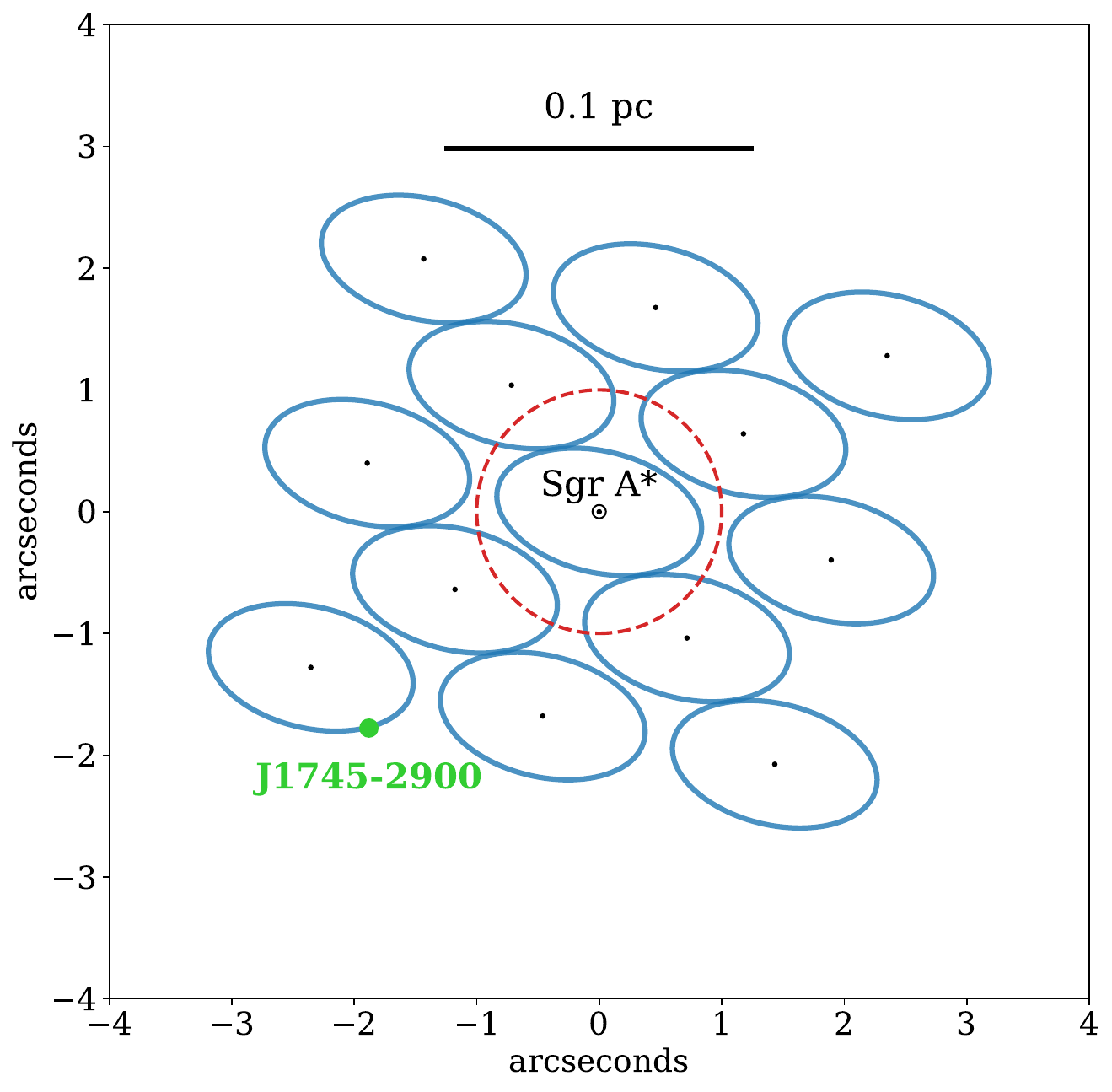}
    \caption{Possible configuration of the beams in a targeted observation around Sgr A*. The center of each of the 16 beams is marked with a black dot. The blue curves show the half-power size of the beams at 9 GHz. For reference we report the position of PSR~J1745$-$2900 in green. The dashed circle shows the area where the S-cluster is located and encompasses all the stars in circular orbit around Srg~A* with orbital period up to $\sim100$ years. All the pulsars useful for the gravity tests described in Section \ref{sec:gravity_tests} would be found in the central beam.}
    \label{fig:SgrA*_beam_tiling}
\end{figure*}

The main target of our searches will be Sgr A* itself. A single tied array beam centered at Sgr A* using all the antennas in the SKA-MID AA4 will have a diameter of $\sim$ 1 arcsec ($\sim 0.04$ pc) and be able to observe pulsars in orbit around the black hole with orbital period less than $\sim 100$ yr. Apart from this beam we want to make use of the beamforming capabilities offered by the SKAO and create an additional 15 beams surrounding the central beam. These beams would allow us to check if the signals detected correspond to a real pulsar. If the signal is seen only in one beam or two beams but not in the others, this means that it is likely not an artifact produced by the receiving system but a real signal localized in the sky. The second major benefit of recording multiple beams is to increase the area covered and maximise the chances of discovery. A possible configuration of the sky positions of the beams is shown in Fig. \ref{fig:SgrA*_beam_tiling}.

The SKA-MID is located in a privileged place to observe the GC. Being located in the Southern hemisphere at a longitude of $-30^{\circ}$S, the GC passes almost at the zenith and is visible above the observing horizon for almost 11 h. We want to take advantage of this by observing the GC for the entire transit guaranteeing us the greatest sensitivity. In order to confirm any candidate found we plan on having several passes of Sgr A*. We plan on observing the target in three different frequency bands: band 5a (5350 - 7850 MHz), the band 5b part 1(8300-10800 MHz) and band 5b part 2 (12900-15400 MHz). We will employ all possible combinations of periodicity search methods.

\begin{figure*}
    \centering
	\includegraphics[scale=0.8]{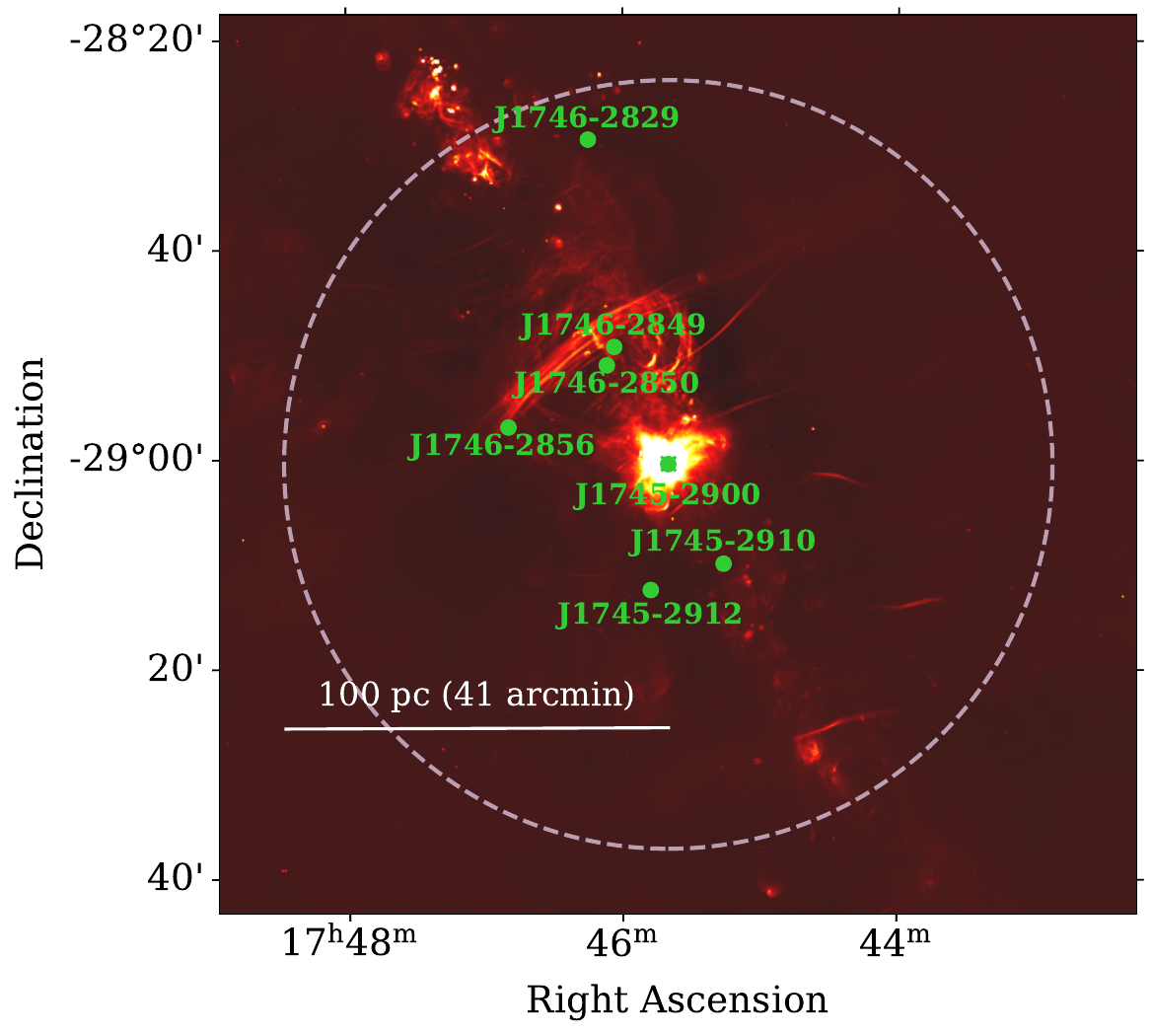}
    \caption{Map of the GC region showing the location of the known pulsars within 100 pc from Sgr A*. The searches are planned over the entire area at band 3 (1650 - 3050 MHz) and band 5a (4600-8500 MHz). Background image from \cite{Heywood+2022}.}
    \label{fig:GC_search_strategies}
\end{figure*}

\subsubsection{Search within the Inner 100 pc}

We also plan on conducting a wider search of the GC up to a radius of 100 pc ($\sim 41$ arcmin). The region is shown in Fig. \ref{fig:GC_search_strategies}. Initially we plan on conducting this search in band 5a (4600 - 8500 MHz). These high frequencies are necessary to counteract the effects of the scattering. The scattering at 4\,GHz as measured from the known pulsars in the region can very from 0.3 ms for PSR~J1746$-$2829 \citep{Wongphechauxsorn+2024} up to $\sim$ 8 ms for PSR~J1745$-$2900 \citep{Eatough+2013}. This suggests that only in band 5a the scattering will be reduced to sub-millisecond levels and the signals of MSPs can be recovered. This region is particularly interesting for pulsars as some may have been dynamically ejected from the central parsec by dynamical interactions with Sgr~A* itself or other objects like stellar mass black holes thought to populate the region \citep{Hailey2018}, informing about the neutron star retention problem. In particular, the search for MSPs will have important consequences on the interpretation of the Fermi $\gamma$-ray excess.

In the future, as the capabilities of SKAO are expanded, we plan on repeating the search also in band 3 (1650 - 3050 MHz) and in band 4 (2800 - 5180 MHz). Due to the typical power-law spectral index, the pulsars are expected to be 3-6 times brighter in these lower frequencies compared to band 5a. The extra flux will boost the signal of slow pulsars increasing the chances of discovery. Surveys at these frequencies will likely not be sensitive to MSPs.

Such a large area can only be covered using the thousands of beams available in the Pulsar Search Pipeline of SKAO. This mode of operation, however, is limited to 300 MHz of bandwidth in AA*.% and only with AA4 will be able to process the entire band. 
The integration time will also be limited to 10-30 minutes. This will allow a complete coverage of the interested area in 5-6 h in band 3 and 20-30 h in band 5a. In order to confirm if the pulsar candidates found in these searches are real, more passes will be necessary.

Apart from a standard acceleration search that will be performed online, we would like to record the raw data in search mode for more advanced searches. These will include all of the techniques described in Section \ref{sec:search_strategies}.

\section{Conclusions}

The pulsar searches with the SKA in the GC region will be the most sensitive searches to date and will have the greatest chances of detecting a pulsar in orbit around Sgr A*. The expected science outcomes from such a pulsar and from other new detections within the GC will enable new valuable astrophysics in several fields. These include tests of gravity theories, studies of the ISM, searches for signatures of Dark Matter, studies of the formation and evolution of the GC and studies of the accretion and feedback of Sgr A*. 

The importance of this dataset is such that new scientific results could be extracted even several years after the actual observations. It can act as in important test-bed for the development of novel search algorithms. This kind of approach has been very successful in the case of observations of pulsars in globular clusters (e.g. \citealt{andersen+2018,Cadelano+18}) where the testing of new algorithms on archival data has led to the discovery new pulsars. For this reason, we desire for the search-mode data of the observations to be archived and eventually be made accessible to the broader pulsar community. The storage of these data would require a few tens of TB for a full transit of Sgr~A*.

%The wealth of information present in these datasets will require time to be fully extracted. While we will make use of state-of-the-art computational facilities and search techniques to process the data to the best of our abilities, we still might miss important discoveries present in the data. For this reason it is important to store the data of as many observations as possible to allow for a deeper reprocessing in the future. This approach has been very successful in globular cluster observations (e.g. \citealt{andersen+2018,Cadelano+18}). The computational power increase predicted in the future and the development of better performing pulsar search techniques can lead to novel discoveries in these observations. \pt{Not sure about this paragraph, as written now it sounds like as we will not be able to properly exploit the data now a days.}

Another important reason to maintain access to these observation is in case future radio, multi-wavelength or multi-messenger observations detect new pulsars in the region. Having knowledge of the position, rotational period, value of DM or orbital parameters will allow us to perform deeper targeted searches that might lead to a re-detection also in the SKA dataset even if the pulsar was not blindly detected at first.

An important contribution in the search of pulsars in the GC will come from the ngVLA \citep{Bower+2018}. It will be able to observe at a wider range of frequencies than the SKA-MID covering from 1.2 to 116 GHz. The observations at frequencies above 15 GHz will increase the chances of discovering MSPs in the case of very strong scattering. At lower frequencies, the larger antennas give the ngVLA the advantage over SKA-MID as far as raw sensitivity but the larger baselines make the ngVLA less suitable to perform surveys over large areas of the sky. The long-baseline astrometric capabilities of the ngVLA will help to accurately measure pulsar positions, proper motions and, potentially, the orbital parameters of pulsars in long-period binary systems. With the advent of SKA2 later on in the future, we will be able to perform even deeper searches potentially discovering a large number of new pulsars. The expected increase in sensitivity of 10 times combined with the improvements in computational power, storage capabilities and performance of the search algorithms is poised to revolutionize our understanding of the GC as a whole.  

% RPE: The detection of more GC pulsars with the SKA can only enhance understanding of this enigmatic region. 

\section*{Acknowledgements}

FA acknowledges that part of the research activities described in this paper were carried out with the contribution of the NextGenerationEU funds within the National Recovery and Resilience Plan (PNRR), Mission 4 – Education and Research, Component 2 – From Research to Business (M4C2), Investment Line 3.1 – Strengthening and creation of Research Infrastructures, Project IR0000034 – ‘STILES -Strengthening the Italian Leadership in ELT and SKA’. ZH, LS, and KJL are supported by the National SKA Program of China (2020SKA0120300,2020SKA0120100), the Beijing Natural Science Foundation (1242018), and the Max Planck Partner Group Program funded by the Max Planck Society. RPE is supported by the Chinese Academy of Sciences President’s International Fellowship Initiative, Grant No. 2021FSM0004. This work is supported by State Key Laboratory of Radio Astronomy and Technology of China. 
%%%%%%%%%%%%%%%%%%%%%%%%%%%%%%%%%%%%%%%%%%%%%%%%
%%%%%%%%%%%%%%%%%%%%%%%%%%%%%%%%%%%%%%%%%%%%%%%%

%\clearpage
%\bibliographystyle{aasjournal}
%\bibliography{Gal_Cen_Pulsars_SKA.bib}
%\vspace{10pt} % OJA format bunched up the generation statement with the references

\bibliographystyle{abbrvnat-maxbibnames4}
\bibliography{Gal_Cen_Pulsars_SKA.bib}

\end{document}